\documentstyle[12pt,aasms4]{article}
\newcommand{\be}{\begin{equation}}
\newcommand{\ee}{\end{equation}}

\def\simlt{\lower.5ex\hbox{\ltsima}}
\def\gtsima{$\; \buildrel > \over \sim \;$}
\def\simgt{\lower.5ex\hbox{\gtsima}}

\def\hmpc{{\rm\,h^{-1} Mpc}}

\def\msunh{{\rm\,h^{-1}~M_\odot}}

\def\ergcm2{\ {\rm erg~cm^{-2} }}
\def\ergscm2{\ {\rm erg~s^{-1}~cm^{-2} }}
\def\hmpc{\ h^{-1}~{\rm Mpc}}

\def\s{\ifmmode \widetilde \else \~\fi}
\def\={\overline}

\def\spose#1{\hbox to 0pt{#1\hss}}

\def\cf{{\it c.f.\ }}
\def\eg{{\it e.g.\ }}
\def\etal{{\it et al.\ }}
\def\ie{{\it i.e.\ }}

\def\lta{\mathrel{\spose{\lower 3pt\hbox{$\mathchar"218$}}
     \raise 2.0pt\hbox{$\mathchar"13C$}}}
\def\gta{\mathrel{\spose{\lower 3pt\hbox{$\mathchar"218$}}
     \raise 2.0pt\hbox{$\mathchar"13E$}}}
\def\mincir{\ \raise -2.truept\hbox{\rlap{\hbox{$\sim$}}\raise5.truept  
\hbox{$<$}\ }}                                                          %
\def\magcir{\ \raise -2.truept\hbox{\rlap{\hbox{$\sim$}}\raise5.truept  %
\hbox{$>$}\ }}                                                          %
\def\simlt{\ \raise -2.truept\hbox{\rlap{\hbox{$\sim$}}\raise5.truept   
\hbox{$<$}\ }}                                                          %
\def\simgt{\ \raise -2.truept\hbox{\rlap{\hbox{$\sim$}}\raise5.truept   %
\hbox{$>$}\ }}                                                          %
\def\newline{\smallskip\noindent}

\def\ea{{\it et al.} \,}

\def\s-z{S-Z}

\lefthead{Colafrancesco at al.}
\righthead{Evolution of the Temperature Function}

\begin{document}

\title{IS THE CLUSTER TEMPERATURE FUNCTION\\
       A RELIABLE TEST FOR $\Omega_0$?}

\author{SERGIO COLAFRANCESCO}
\centerline{Osservatorio Astronomico di Roma}
\centerline{Via dell'Osservatorio 2, I-00040 Monteporzio, Italy \\}
\author{PASQUALE MAZZOTTA}
\centerline{Dipartimento di Fisica,  Universit\`a di Roma ``Tor Vergata''} 
\centerline{Via della Ricerca Scientifica 1, I-00133 Roma, Italy\\}
\and
\author{NICOLA VITTORIO}
\centerline{Dipartimento di Fisica,  Universit\`a di Roma ``Tor Vergata''} 
\centerline{Via della Ricerca Scientifica 1, I-00133 Roma, Italy\\}

\vspace{.2in}


\begin {abstract}
We discuss the evolution of the cluster temperature function (TF) in different 
scenarios for structure formation.
We use the commonly adopted procedure of fitting the model 
parameters to the local TF data, finding the best fit values and, most of all, 
 the associated statistical errors.
These errors yield an uncertainty in the prediction of the TF evolution. 
We conclude that, at the moment,  
observations of cluster temperatures at $z \simlt 0.5$
could provide only a weak test for $\Omega_0$.
\end{abstract}


\section{Introduction}
The gravitation instability in a flat universe  seems to provide a
reasonable scenario for structure formation. COBE/DMR (Smoot \etal
1992; Gorski \etal 1994; Banday \etal 1994)  and the more recent CMB
anisotropy experiments at degree angular scales (see \eg de Bernardis
\etal 1996 for a discussion) unvealed the presence of linear density
fuctuations at recombination, in quite a good agreement with the
theoretical predictions of dark matter dominated models. However, the
non-linear evolution of these  fluctuations and the actual formation of
objects like galaxies or clusters of galaxies is far from being
understood. Being  the latest structures formed by the present time via
a dissipationless collapse, clusters of galaxies play a crucial role in
linking the linear and non-linear regimes of the gravitational
instability theory. So, from one hand  the Press \& Schechter (1974,
hereafter P\&S; see also Bond \etal 1991) theory allows to predict the
abundance and evolution of clusters of galaxies in a simple and
semi-analytical way, by using the linear theory of structure formation. 
On the other hand, the available surveys of clusters of galaxies in the
X-rays allow to construct observables like the local luminosity  
(see, \eg, Kowalski et al. 1984 and Ebeling \ea 1997), 
and temperature 
(Henry \& Arnaud 1991, hereafter H\&A; Edge et al. 1990) 
functions. 

Theoretical predictions for the X-Ray Luminosity Function (XRLF) heavily
rely upon assumptions on the amount, the distribution and the evolution
of the intra-cluster (IC) gas. Observations seem to indicate that the
IC gas is a fraction between 10 and perhaps 30\% of the cluster virial
mass (White \etal 1993, Cirimele, Nesci \& Trevese 1997). 
Moreover, even for virialized clusters the
physics behind the formation of a core in the gas distribution is not
known. Finally, there have been claims that the IC gas rapidely decrease 
for redshifts larger than $0.3$ (Gioia \ea 1990, Henry \ea 1992, Cavaliere \&
Colafrancesco 1988, Kaiser 1991), although  it is very hard at the moment
to draw definitive conclusions on this issue. 
In fact, it seems that the amount of XRLF evolution
at $z\simlt 0.5$ is somewhat less 
(Ebeling \ea 1997, Nichol \ea 1996, Romer \ea 1997)
than suggested by the
Extended Medium Sensitivity Survey (Gioia \ea 1990, Henry \ea 1992).
Because of  these
uncertainties, it has been widely argued that the cluster temperature
function (TF) is a more reasonable quantity to work with (H\&A, Kaiser 1991). 
The basic
point is that the temperature of the  IC gas in hydrostatic equilibrium
with the potential well of a virialized cluster depends only on the
cluster virial mass.  

In spite of this simplification, it turns out that in dark matter, 
COBE/DMR normalized models the theoretical TF is not consistent with the
local data (Eke \etal 1996, hereafter Eke \ea). 
Thus, the usual procedure is to tune the
model parameters to fit  the local TF  and, then, to make predictions
for other observables such as the cluster number counts and the
Sunyaev-Zel'dovich  effect (Barbosa \ea 1996, Eke \ea). 

Following this procedure, it has been shown that: {\it i)} the TF evolution in
flat and open cosmologies is drastically different;
{\it ii)} observations
of cluster temperatures at redshifts $z \simgt 0.3$ can strongly constrain
$\Omega_0$ (Hattori and Matsuzawa 1995, hereafter H\&M;
 Eke \etal
1996, Barbosa \etal 1996, Oukbir and Blanchard 1992, Oukbir and  
Blanchard 1997). 
While we fully agree with point {\it i)}, we believe that point {\it ii)}
requires a more carefull investigation.

Thus, the purpose
of this paper is to show that: {\it i)}  the
amount of TF evolution is heavily determined by the fit  to the local
data, and is not a self-consistent prediction of the theoretical models;
{\it ii)}  the intrinsic statistical uncertainty of this fit
smeares out the difference between the low density and $\Omega_0=1$
model predictions for the TF evolution. 

The plan of the paper is as follows. In  Sect.2 we
review the basic steps behind  the theoretical calculations.
 In Sect.3  we discuss in a simple semi-analytical way what determines
the  TF evolution. In Sect.4 we show the constraints on the model
parameters set by the fit  to the local TF. In Sect.5 we present
detailed numerical predictions of the TF evolution in different
structure formation models, including Cold Dark Matter (hereafter CDM)
cosmologies (flat, open and
vacuum dominated). Finally, in Sect.6 we summarize our main findings.

\section{The cluster TF}
The  cluster TF is defined as
\be
N(T,z) = {\cal N}(M,z) {dM\over dT}
\ee
where  ${\cal N}(M,z)$ is the cluster mass function (MF). The  latter is
usually derived by the  P\&S  theory:
\be
{\cal N}(M,z) = \sqrt{2\over\pi} {\rho\over M^2} {\delta_v \over
\sigma}{dln
\sigma 
\over dln M}\exp[-\delta_v^2/2\sigma^2] \, ,
\ee
where $\rho$ is the comoving background density, $M$ is the
total cluster mass and 
$\delta_v$ is the linear density contrast of a perturbation that
virializes at $z$. The  variance of
the (linear) density  fluctuation field at the scale  $R=(3 M / 4\pi
\rho)^{1/3}$ and redshift $z$  is given by the standard relation
(see \eg Peebles 1980):
\be
\sigma^2(R,z)= {D^2(\Omega_0,z) \over 2 \pi^2} 
\int k^2 dk P(k) \bigg[{3 j_1(kR) \over kR}\bigg]^2
\ee
where $D(\Omega_0,z)$ is the growth factor of linear density fluctuations in a
given cosmology, and $j_1$ is a spherical Bessel function. 
If we normalize the matter power spectrum, $P(k)$, by requiring
$\sigma(8 \hmpc,0) = b^{-1}$, then the MF depends on the product $b
\delta_v$,  and not separately on $\delta_v$ and $b$.

While the cluster MF gives the mass and redshift distribution of a 
population of evolving clusters, the Jacobian $dM/dT$ describes the 
physical properties of the single cluster. Under the standard assumption
of the IC gas in hydrostatic equilibrium with the potential well of a
spherically simmetric, virialized cluster, the IC gas temperature--cluster 
mass relation is easily obtained by applying the virial
theorem: $T=-\mu m_p U/(3K_B M)$, where $\mu=0.62$ is the mean molecular 
weight (corresponding to a Hydrogen mass fraction of $0.69$), 
$m_p$ is the proton mass,
$K_B$ is the Boltzmann constant and $U$ is the cluster potential energy. 
If the cluster is
assumed to be uniform, $U=-(3/5)GM^2/R_v$ and $T=T^{(u)}\equiv (1/5) (\mu m_p/
K_B)GM/R_v$, where $R_v= [3M/(4\pi \rho \Delta)]^{1/3}/(1+z)$ is the
cluster virial radius and
$\Delta(\Omega_0,z)=18\pi^2/[\Omega_0(H_0t)^2(1+z)^3]$ is the non
linear density contrast of a cluster that virializes at $z$ in a
$\Omega_0\le 1$ cosmological model [in flat, vacuum dominated low
density models $\Delta$ has not an analytical expression (see, \eg
Colafrancesco \etal 1997 and references therein)]. 

We also relax the assumption of
uniformity by considering a 3-D gas density profile:
\be
n(r) = n_c \bigg[ 1 + \bigg({r \over r_c} \bigg)^2 \bigg]^{-3\beta/2}
\ee
where $n_c$ is the central electron density and $r_c$ is a core 
radius. 
The mass within the outer (proper) radius, taken as $R=pr_c$, is
$
M(p) = 3 M_c \omega(p, \beta),
$
where $\omega(p,\beta)= \int_0^p t^2dt/(1+t^2)^{3 \beta/2}$,
$t\equiv r/r_c$,
$M_c=(4\pi/3)r_c^3\rho_c$, and $\rho_c$
is the central total mass density of the cluster. Because of the
assumed profile, the ratio between the central and mean mass density
of the cluster is $\rho_c/(\rho \Delta)=p^3/3\omega(p,\beta)$. Then,
$r_c=R_v/p$, or, equivalently, 
\be
r_c(\Omega_0,M,z) = {1.29 \hmpc \over p} \biggl[ {M\over M_{15}}
\cdot {\Delta(1,0)\over \Omega_0 \Delta(\Omega_0,z)}\biggr]^{1/3}
{1\over 1+z}   
\ee
where $M_{15}=10^{15} \msunh$.
In this case, $U=-(GM^2/r_c) \psi/\omega^2$ and $T= 5 p \psi
T^{(u)}/(3\omega^2)$, where $\psi(p,\beta)=\int_0^p s
ds/(1+s^2)^{3 \beta/2} \int_0^st^2dt/(1+t^2)^{3 \beta/2}$. Hereafter we fix
$p=10$ to recover $r_c$ values consistent with 
the observations (see \eg Henriksen and Mushotsky 1985,
Jones and Forman, 1992). 

At this point, the cluster TF is fully determined as we know
\be
M(T,z) \propto {T^{3/2} \over \sqrt{\Omega_0 \Delta_v(\Omega_0,z)} }
{1 \over (1+z)^{3/2}} {1 \over h}
\ee
and $dM/dT=3M/2T$.

\section{The TF  Evolution: an analytical approach}
In this Section we restrict ourselves to the simple case of power--law
power spectra (hereafter PLPS). This allows us to write down an explicit
expression for the TF, by using the results of Sect.2. In particular,
the $M(T,t)$ relation provides for the mass variance (in this case
$\propto M^{-\alpha}$) the following expression:
\be
\sigma[M(T,t),t]={1\over b}
\biggl({T\over T_0}\biggr)^{-3\alpha/2} \biggl({t\over
t_0}\biggr)^{-\alpha} D(\Omega_0,t)
\ee
where $\alpha=(n+3)/6$ and $n$ is the spectral index. Here $T_0$ is the
present IC gas temperature of a cluster of mass $M_0$, corresponding to 
the normalization scale of $8h^{-1}Mpc$. Because of the use of $T$ and
$t$ as independent variables, the time dependence of $\sigma$ is
different from the standard one: the extra factor, $(t/t_0)^{-\alpha}$,
takes into account the  fact that a cluster  with a given temperature at
$t<t_0$ is
less massive than  a cluster with the same temperature at
$t_0$. Along the same line, it is easy to verify that:
\be
{1\over M^2} {dM\over dT} = {3\over 2} {1\over M_0T_0} \biggl({T\over
T_0}\biggr)^{-5/2} \biggl({t\over t_0}\biggr)^{-1}
\ee
Equations (7) and (8) allow to write the TF as follows:
\be
N(T,z)=C
 \biggl({T\over
T_0}\biggr)^{(-5+3\alpha)/2} \biggl({t\over
t_0}\biggr)^{-1+\alpha}{1\over D} \exp
\biggl[-{(\delta_vb)^2\over 2}\biggl({T\over
T_0}\biggr)^{3\alpha} \biggl({t\over
t_0}\biggr)^{2\alpha}{1\over D^2}\biggr]
\ee
where $C$ is a constant which depends on $\Omega_0$, $h^3$, and
$\delta_vb$. For didactic purposes, it is worth  to derivate Eq.(9)
w.r.t. time: this will help to understand which quantities define the
amount of TF evolution. It is immediate to verify that:
\be
{dlnN\over dt}=-{1\over t} - {1\over \tau_-} + {1\over \tau_+} 
\ee
where $1/\tau_- = \dot D/D-\alpha/t$ and $1/\tau_+ =
(\delta_vb)^2(T/T_0)^{3\alpha} (t/t_0)^{2\alpha}/(D^2 \tau_-)$.

The amount and kind (positive or negative) of TF evolution is set by
the competition of these three time scales: the cosmic time $t$, $\tau_-$ and
$\tau_+$. When either $t$ and/or $\tau_-$ are shorter than $\tau_+$,
then the TF evolution is negative. When $\tau_+$ is shorter than 
either $t$ or $\tau_-$, then the TF evolution is positive. At a given
cosmic time, $\tau_-$ depends on the chosen cosmology through the
logarithmic derivative of the growth factor. In principle $\tau_-$
depends also on the shape of the power spectrum: in practise, for
$\Omega_0\le 1$ models with no cosmological costant, $\alpha/t < \dot
D/D$. The other time scale, $\tau_+$, at a given time and temperature
depends weakly on the spectral index $n$ and quadratically on the
$\delta_vb$ parameter. We call the reader's attention on this point. In
fact, at least for PLPS, both $n$ and $\delta_vb$ are determined  by
fitting the theoretical prediction for the local TF to the existing
data. Thus, for a given cosmology, the TF evolution can be positive or
negative, depending on the derived values for $n$ and, most of all, for 
$\delta_v b$.
So, while fitting the theoretical
TF to its local, observed values seems "the best one can do" in the
absence of a more refined theory for cluster formation and evolution,
we can not forget that the amount and kind of the TF evolution is
already  determined just by this fitting procedure.

Similar conclusions are of course  reached if we use redshift instead of
cosmic time as an independent variable. In this case we have
\be
{dN(T,z)\over dz} = -{dt\over dz} N(T,z) {d\ln N(T,t)\over dt}
\ee
where, we remind it, $N(T,t)$ is itself proportional to the
$\delta_vb$, while ${dt/dz}$ depends only on the chosen cosmology.

The purpose of this Section was to show qualitatively the central role
that $\delta_vb$ plays (together with $n$, at least in the PLPS models) 
in defining the TF
evolution for a given cosmological model. Thus, we believe that it is
important not only to estimate  best fit values, but also the
statistical uncertainties associated with them. This is the goal of the next Section.

\section{Fit to the local FT}
In this Section we
derive some model parameters by fitting the TF theoretical predictions
to the  H\&A data. 
%
%
We use flat and open models where density fluctuations are described 
by PLPS,  and CDM cosmologies, either flat  or open  or vacuum
dominated.

We use the following mass-temperature
relation [see Eq.(6)]:
\be
M=M_{15}
\biggl({T\over T_{15}}\biggr)^{3/2} \biggl({t\over
t_0}\biggr) 
\ee
Accordingly to the uniform cluster
model described in Sect.2, a cluster of mass $M_{15}$
has a temperature $T_{15}=4.3$ keV.
However, to compare our results with those of 
H\&A and H\&M we will use in the following $T_{15}=6.4$ keV.
This value was choosen by H\&A and H\&M to be consistent with
the numerical results of Evrard (1991, hereafter Evrard), who also 
found from his 
simulations a scatter of $\approx 10\%$ in the M-T relation. 

Consistently with the observations, we express
the TF  as the number of cluster per $(h^{-1}Mpc)^3$ per
keV. 
Then, the  PLPS model predictions for the TF are independent of
$h$ [cf. Eq.(9)]. 
Thus, we consider as free parameters both
$\delta_vb$ and the spectral index $n$. 
The best fit values are shown in Table 1, together with the uncertainty in 
the fit at the 68.3\% and, in parenthesis, at the 90\% confidence level.
%
%
%
%
Our uncertainties 
are smaller than those quoted by H\&A. This is
because we did not consider (as H\&A did) the $10\%$
uncertainty on the $M-T$ relation derived from 
Evrard  simulations. However,  even at the level we quoted them,  
these uncertainties  have serious impact on the
model predictions as we will show in the next Section.

Note that $\delta_v=2.2$ is the density contrast at 
virialization extrapolated from the linear theory.
This numerical value changes very little varying $\Omega_0$ and $z$.
If we assume that all clusters are virialized, then the value of the 
biasing parameter is of the order of unity for $\Omega_0=1$,
but less than unity (antibias) for low density universes.
This is due to the following fact.
The mass contained in a $8\hmpc$ sphere is $M_0=0.63 \Omega_0 
M_{15}$.
According to Eq.(12), the collapse of such a 
sphere will results in clusters with temperature of $4.7$,
$1.6$ and $1.0$ keV
for $\Omega_0=1$, $0.2$ and $0.1$,  respectively.
In order to form Coma-like clusters   
(\ie with  $M \simeq   M_{15}$ and
$T\simeq 6$keV) in a low density universe, 
we have to assume that those density fluctuations that generate 
clusters have amplitudes larger than the mass density field:
in other words, clusters must be antibiased.
Our values of $b$ are quite consistent with the
scaling found by White \etal (1993):
%
%
$b=1/\sigma(8 \hmpc) \simeq 1.75\Omega_0^{0.56}$.
In fact, we find $b =1.7 \Omega_0^{0.58}$ if we choose $\delta_v=1.68$
(if we identify clusters as collapsed rather than virialized objects).
The small difference between our result and that of White \ea (1993) depends 
on the different datasets used for the cluster abundance.

For CDM cosmologies, we use both the uniform cluster  and the
$\beta$-profile models.  If we use the
uniform cluster model, with  fixed $\Omega_0$ and $h$, we get the same
results as Bartlett and Silk (1993).
If we assume a  $\beta$-profile for the IC gas distribution, then 
$T_{15}=5.8$ keV and $7.8$ keV for $\beta=2/3$ and $0.93$, respectively. 
The latter temperature value is the one found by 
Eke \ea analyzing numerical
simulations, which they show to be quite consistent with the
P\&S predictions. 

At variance with the PLPS models, in the CDM scenario $h$ 
contributes, together  with $\Omega_0$, to define the shape of the power
spectrum. So, keeping the $\beta$-profile for the IC gas
distribution with $\beta=2/3$ and assuming $n=1$, 
we fit the local FT data by considering
as free parameters $\delta_vb$, $\Omega_0$  and $h$.
We get the following best fit values:
$\Omega_0=0.4^{+0.10(+0.15)}_{-0.11(-0.14)}$,
$h=0.6^{+0.4(+0.4)}_{-0.1(-0.1)}$ and
$\delta_vb=1.6^{+0.05(+0.14)}_{-0.13(-0.19)}$. 
Note that we force $h$ to be in the range: $0.5 \leq h \leq 1$.

Although a reasonable fit is obtained in a quite broad region of the
parameter space, it must be stressed that,  as already
found by H\&A, the standard flat CDM model
with $h=0.5$ does not provide a good fit to the TF data: 
for such a model the reduced $\chi^2$ is $1.8$ (see Table 2). 
Our results for CDM models are also consistent, within the quoted
uncertainties, with those found by Viana and Liddle (1996), using
$N(7 {\rm keV},z=0)$ to derive the amplitude of the
fluctuation spectrum.
The corresponding uncertainties are of the order of $+35\%$ and
$-25\%$ at 95\% confidence level for low density, vacuum dominated CDM  
models (hereafter CDM+$\Lambda$), quite independently on the assumed
modelling of the TF.

%
%
%

\section{Results}
Now we can use the best fit values to predict the cluster TF evolution.
Let us first consider PLPS models. With the best fit values of $n$ and
$\delta_vb$, we find that the TF evolution is quite
different in  open and flat models: the abundance of clusters of a
given temperature predicted in an $\Omega_0 =0.1$ universe
 is  substantially larger than in the flat model, a result already
obtained by  
H\&M. However, this does not necessarely imply
that we can estimate $\Omega_0$ by using data
on the TF evolution. We have first to quantify the
probability of concluding that the universe is low (high) density when it is
actually high (low) density. 
Let us assume that the TF depends only on $b$. Then, the uncertainty 
on $N(T,z)$ due to the uncertainty on
$b$ (derived from fitting to the local TF) writes as:
\be
{\Delta N \over N} = {1\over N} \biggl|{\partial N\over \partial b}\biggr| \Delta
b
\ee
The TF can be written as
$
N(T,z)\propto g \exp[-0.5g^2]
$
where for PLPS
$
g=[\delta_vb/D(t)] (M/ M_0)^\alpha 
$.
Because of our mass-temperature relation [\cf Eq.(12)] we can write
(for $\Omega_0\leq 1$ and $\Lambda = 0$)
\be
g= {\delta_vb \over D(t)} \biggl({T\over T_0}\biggr)^{3\alpha/2 } 
\biggl({t\over
t_0}\biggr)^{\alpha }
\ee
and, if  we further assume $\Omega_0=1$,  
\be
g= \delta_vb (1+z)^{1-3\alpha/2} \biggl({T\over T_0}\biggr)^{3\alpha/2 }
\ee
So we have:
\be
\biggl({\Delta N\over N}\biggr)=\biggl({\Delta b\over b} \biggr)
\times \biggl|\biggl[1- (\delta_vb)^2 (1+z)^{2-3\alpha}
\biggl({T\over T_0}\biggr)^{3\alpha}\biggr]\biggr|
\ee
Now, for $\Omega_0=1$ and PLPS, the best fit values are 
$n=-1.8$  and $\delta_vb=2.8$, the latter with an uncertainty 
$\Delta b/b \simeq 9\%$ at the 90\% confidence level. 
This uncertainty yields an uncertainty on
$N(T,z)$ which increases either  with 
$T$ and/or with
$z$,  as $2>3\alpha$. Moreover, in an $\Omega_0=1$ universe, clusters with
$M=M_0$  have a temperature $T_0=6.4 (0.63)^{2/3}$keV. As a
result, for $\Omega_0=1$ and $n=-1.8$ we get:
\be
\begin{array}{lr}
{\Delta N\over N}(T, z=0.0)  &= \biggl({\Delta b\over b} \biggr) \times
\biggl|[1-  3.1 T^{0.6}(keV)]\biggr| \\
{\Delta N\over N}(T, z=0.5)  &= \biggl({\Delta b\over b} \biggr) \times
\biggl|[1-  5.5 T^{0.6}(keV)]\biggr| \\
{\Delta N\over N}(T, z=1.0)  &= \biggl({\Delta b\over b} \biggr) \times
\biggl|[1-  8.2 T^{0.6}(keV)]\biggr| \\
\end{array}
\ee
From Eq.(17) we found  that,  at the $90\%$ confidence level, 
$\Delta N/N
\simgt 1$ (\ie more than 
$100\%$ uncertainty!) for
$T\simgt 9.7$,  $3.7$ and $1.9$keV at
$z=0$, $0.5$, and $1$, respectively. 

We plot in Fig.1 the 
region spanned by the predicted TF, for both
$\Omega_0=1$ and $\Omega_0=0.2$, once the statistical uncertainty 
(at the $90 \%$ c.l.) on the best fit values for both $n$ and
$\delta_vb$ are taken into account. 
Thus, if we write
the TF as $N(T,z|\sigma_8,n)$, the 
uncertainty region is bounded by 
$N(T,z|\sigma_8+\Delta\sigma_8,n-\Delta n)$ and
$N(T,z|\sigma_8-\Delta\sigma_8,n+\Delta n)$. Here
$\Delta\sigma_8$ and $\Delta n$ are the uncertainties on
$\sigma_8$ and $n$ derived,  at a given confidence level,  
form the fit to the local TF (see Table 1).  

It is evident the large degree of
overlap of the two regions in the $3\div 10$ keV range, up to redshift of
order of unity. This overlap would have been even more substantial if
we had taken into account the
$10\%$ uncertainty in the
$M-T$ relation implied by the Evrard simulations.
This is why we believe that the 
H\&M conclusion that the TF evolution  can test,
at least for PLPS models and in the framework of the described
procedure, the geometry of the universe has to be taken with care. 
As found by Eke \etal, the TF evolution is basically the same
in low density, open and vacuum dominated models up to  $z \approx 0.5$.

The  behaviour of the uncertainty region  can be completely understood
with the help  of Eq.(17) above. There is a temperature such that
$\Delta N/N=0$ even if $\Delta b/b\neq 0$: this explains the region 
of minimum
(or even zero) uncertainty, which depends upon $\Omega_0$, $z$, 
and $\delta_vb$.
Obviously, the region of maximum uncertainty is around the cutoff 
temperature, $T_c$:
small variations in the
parameters  result in huge TF variations at
$T\simgt T_c$.  
The uncertainty region shown in Fig.1 
spans several orders of magnitude: this is simply an artifact of the
logarithmic scale: having uncertainties of $\approx
100\%$ obviously means 
$0\simlt N(T,z) \simlt 2 {\overline N}$, where ${\overline N}$ is the 
TF obtained with the best fit values.

We reach similar conclusions for the CDM scenario. 
With the best fit values for $h$ and $\delta_vb$ (at given
$\Omega_0$; see Table 2), we conclude that the TF evolution is 
again quite different
in low-density ($\Omega_0=0.3$, vacuum  dominated, say) and flat models.
The former is the same cosmological model considered by Eke \etal
However, the
$90\%$ confidence level regions obtained by considering the statistical
uncertainties on the best fit values are substantially overlapped in the
$3\div 10$ keV up to
$z\sim 0.5$ (see Fig.2).  
The uncertainty  region in Fig.2 is evaluated as in the PLPS
case. 
Moreover, as the flat CDM model does not provide a reasonable fit to 
the local TF, we show the uncertainty regions for two CDM$+\Lambda$ 
models, with $\Omega_0=0.7$ and $0.3$, respectively.
We obtain very similar
results for the cluster redshift distributions (see Fig.3).


So, from one hand we agree with the conclusion of Eke \etal that
"even at $z=0.33$, 
these (temperature) distributions depend very strongly on $\Omega_0$''.
On the other hand, normalizing to the local TF yields large
uncertainties in the theoretical predictions of the cluster TF evolution.
Because of these uncertainties, we do not believe that measurements of the
cluster TF at high $z$ can yet provide a good estimate  for $\Omega_0$.
We want also to stress that assuming $\beta\simeq 1$, only to reproduce
the Eke \etal $T_{15}$ value, would have increased the degree of overlap
of the 90 \% c.l. regions of Fig.2 even at $z\simeq 1$.

\section{Discussion and conclusions}
We have shown that the present data on the local TF and the lack of a
self-consistent theory for cluster formation and evolution strongly
weaken the predictive power of the theory. 
In fact, as discussed in Section 3, normalizing
the models parameters (mainly $\delta_vb$) to the local TF
determines the degree and kind of the
TF evolution. 
Moreover, just because of the quality
of the available data, the fit can not be so precise to have a clear
cut distinction between the predictions of low-density and flat
models. This is why we believe that, although in principle possible,
any test of the geometry of the universe using the cluster TF at
intermediate redshifts ($z\simlt 0.5$) can not be, at the moment, very 
precise. 
We want to stress that our quoted uncertainties on the best fit values
are systematically lower than those quoted by other authors.
The uncertainty regions of Figs. 1 and 2 would have been broader if we 
had: {\it i)} included a 10\% scatter in the $M-T$ relation derived 
from the Evrard simulations;
{\it ii)} used the best fit model, \ie CDM$+\Lambda$ with 
$\Omega_0=0.4$; {\it iii)} included other systematic uncertainties 
related to measurements of the cluster temperatures and to the 
catalogue incompleteness (see Eke \ea); {\it iv)} considered for $h$ 
the actual uncertainties derived from the $\chi^2$ analysis, without 
limiting the allowed $h$ values to the standard interval $0.5 \leq h \leq 
1$; {\it v)} used the Eke \ea value of $T_{15}=7.8$ keV,
instead of our $T_{15}=5.8$ keV.

In any case, even if no uncertainties were included (\ie all 
the parameters such as $b \delta_v$ and the others were known with very high
precision) we find that
in order to reject the hypothesis of an $\Omega_0=1$, CDM  
universe (at $90 \%$ confidence level), we should
observe  $\simgt 16$, $7$ keV clusters per steradian at $z\approx 0.3$.
The intrinsic statistical uncertainty of the $N(T=7 {\rm keV}, z =0.3)$
would be $\approx 2 \times \sqrt{N}$.
Such a measurement would then provide 
$\Omega_0 \approx 0.7^{+0.3}_{-0.2}$, again, at the  $90 \%$ confidence level.
However, if one takes into accont the parameter uncertainties, this estimate 
weakens quite considerably (see Fig.4).
In any case, in order to have a complete sample of high-$z$ clusters with 
precise
temperature determinations we have to wait for the next generation of
X-ray space experiments, featuring good energy resolution ($E/\Delta E \simgt
50$), large photon collecting areas $A_{eff} \simgt 2 \times 10^2 cm^2 
(@ 7 keV)$
and a large sky coverage ($\simgt$ a few sr).
The planned X-ray missions of the next decade (such as AXAF, XMM, 
SPECTRUM-X-$\gamma$, ABRIXAS, ASTRO-E and HTSX) could yield such detailed
informations on high-$z$ clusters 
provided that a total observing time $\simgt$ a few $ 10^6 s$ 
will be devoted to such studies.

\acknowledgements{{\bf Acknowledgements} 
We are indebted with the Referee for helpful comments and 
suggestions that improved the presentation of our results.}

\newpage


\newpage 


\figcaption{
Uncertainty regions  in the TF predictions for
PLPS models at redshifts $z=0.3$ (upper
panel), $z=0.5$ (middle panel) and $z=1$ (lower panel).
The shaded (dashed for $\Omega_0=1$ and dotted for $\Omega_0=0.2$)
areas are drawn considering the uncertainties (at $90\%$ confidence 
level) 
in the model parameters (see text).
}

\figcaption{
Same as Figure 1 but for two low-density,
vacuum dominated CDM models: $\Omega_0=0.3$ (dotted shaded area) and 
$\Omega_0=0.7$ (dashed shaded area).
}

\figcaption{
Uncertainty regions for the redshift distributions of clusters with 
$T > 3 keV$ (panel a) and $T > 7 keV$ (panel b) as 
predicted by the same models of Figure 2.
}

\figcaption{
Cluster TF for the flat CDM model ($\Omega_0=1$, $h=0.5$,
$\delta_v b =2.5$: continuous line) and for two CDM$+\Lambda$ 
models ($\Omega_0=0.7$, $h=0.5$,
$\delta_v b =2.05$: dashed line; $\Omega_0=0.5$, $h=0.6$,
$\delta_v b =1.78$: dot-dashed line).
The shaded area shows the uncertainty region for a low density 
($\Omega_0=0.3$) CDM$+\Lambda$) model, once the 
uncertainties on $\delta_v b$ and $h$ are taken into account.
}

\newpage

\medskip
\begin{center}
{\bf Table 1.} Fitting parameters for the TF: PLPS models.
\end{center}
\begin{center}
\begin{tabular}{|l|l|l|l|}
\hline
$\Omega_0$ & $n$ & $b \delta_v$ & $\chi^2_{min}$ \\ 
\hline
& & &  \\
1   & $-1.8^{+0.41(+0.75)}_{-0.44(-0.90)}$    & 
$2.8^{+0.17(+0.24)}_{-0.10(-0.25)}$          &  0.66 \\
& & &  \\
0.2 & $-0.9^{+0.63(+1.25)}_{-0.72(-1.13)}$    &
$1.1^{+0.23(+0.38)}_{-0.23(-0.41)}$          &  0.59  \\     
& & &  \\
0.1 & $-0.5^{+0.53(+1.19)}_{-0.44(-1.00)}$   &
$0.6^{+0.11(+0.31)}_{-0.13(-0.27)}$          &  0.74 \\
& & & \\
\hline
\end{tabular}
\end{center}

%

\medskip
\begin{center}
{\bf Table 2.} Fitting parameters for the TF: CDM models.
\end{center}
\begin{center}
\begin{tabular}{|l|l|l|l|l|}
\hline
Model & $\Omega_0$ & $b \delta_v$ & $h$ & $\chi^2_{min}$ \\ 
\hline
& & & &  \\
SCDM & 1 & $2.5^{+0.0(+0.03)}_{-0.0(-0.03)}$   & 
$0.5^{+0.0(+0.02)}_{-0.0(-0.0)}$    & 1.84 \\
& & & &  \\
CDM$+\Lambda$ & 0.4 & $1.6^{+0.05(+0.14)}_{-0.13(-0.19)}$ & 
$0.6^{+0.4(+0.4)}_{-0.1(-0.1)}$ & 0.58 \\
& & & &  \\
CDM$+\Lambda$ & 0.3 & $1.4^{+0.11(+0.23)}_{-0.19(-0.25)}$ &  
$0.7^{+0.3(+0.3)}_{-0.2(-0.2)}$ & 0.74 \\
& & & & \\
CDM & 0.3 & $1.5^{+0.10(+0.23)}_{-0.20(-0.25)}$  & 
$0.65^{+0.35(+0.35)}_{-0.15(-0.15)}$ & 0.72 \\
& & & &  \\
\hline
\end{tabular}
\end{center}
\noindent
%
%
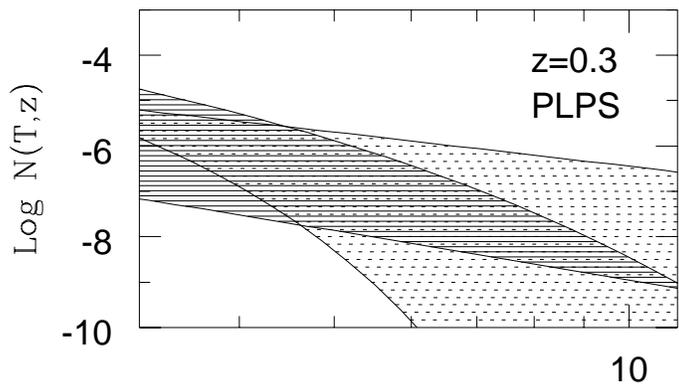
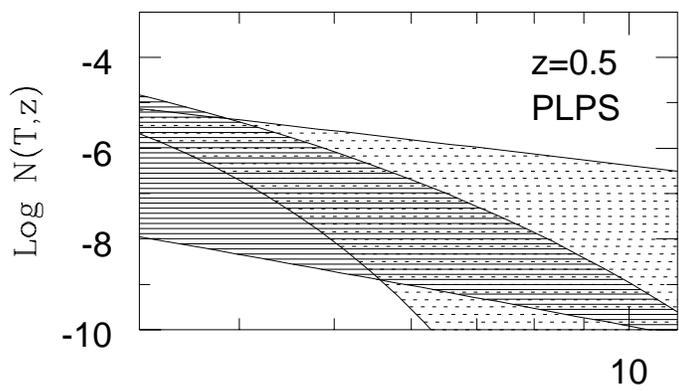
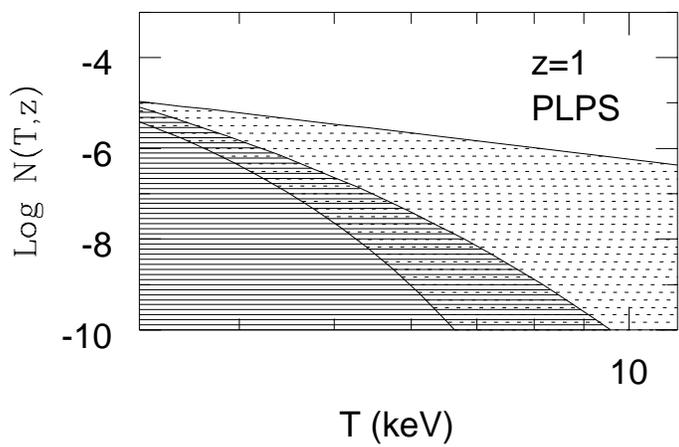
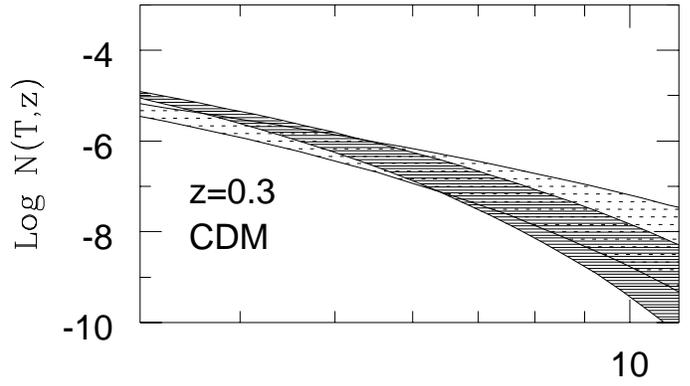
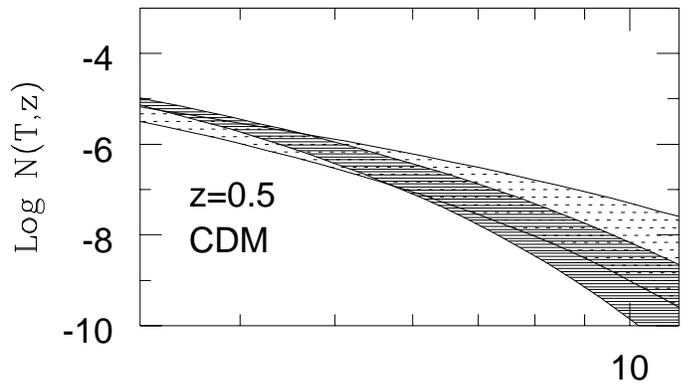
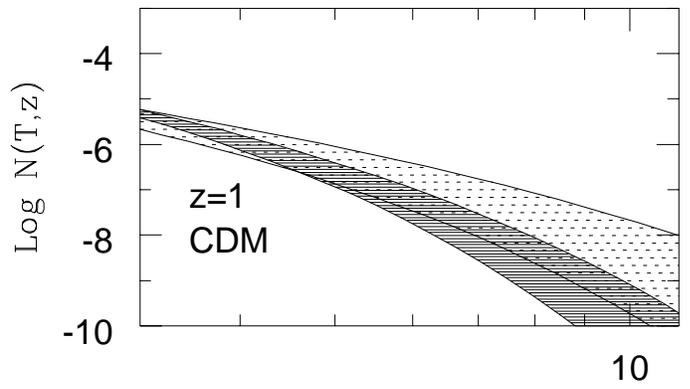
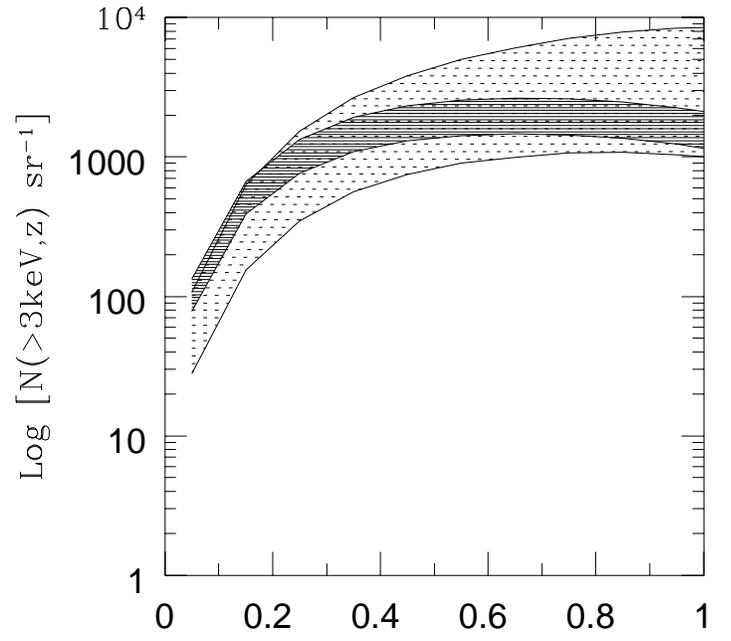
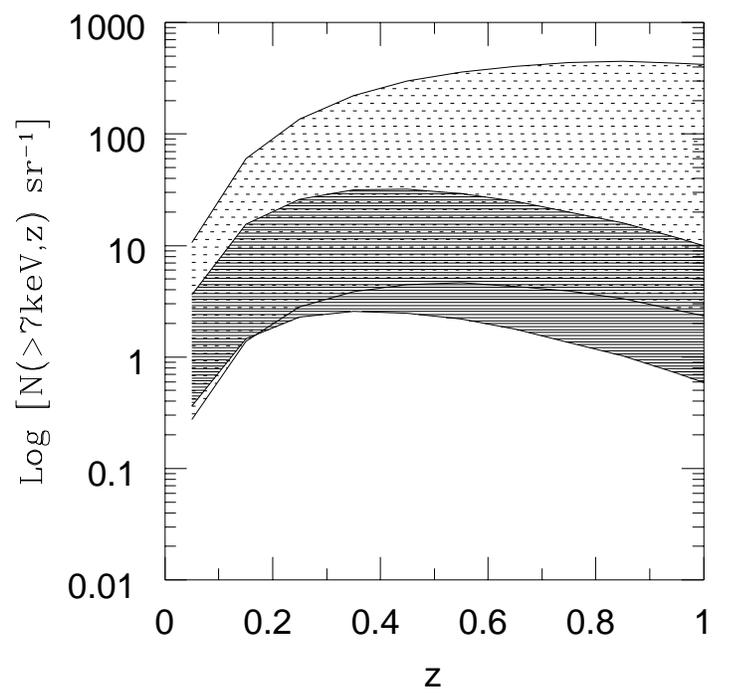
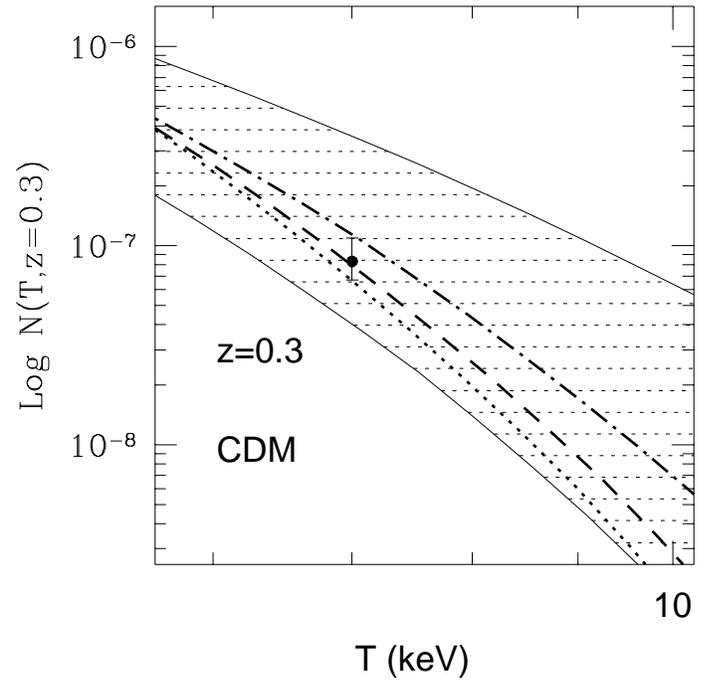
\end{document}